# Probing the Thermal Deoxygenation of Graphene Oxide using High Resolution *In Situ* X-Ray based Spectroscopies

*Abhijit Ganguly,[§] Surbhi Sharma,[§] Pagona Papakonstantinou*, and Jeremy Hamilton*

Nanotechnology and Advanced Materials Research Institute, NAMRI, University of Ulster, Jordanstown campus, BT37 0QB, United Kingdom.

[§]These authors contributed equally.

*Corresponding author, E-mail: p.papakonstantinou@ulster.ac.uk

**ABSTRACT.** Despite the recent developments in Graphene Oxide due to its importance as a host precursor of *Graphene*, the detailed electronic structure and its evolution during the thermal reduction remain largely unknown, hindering its potential applications. We show that a combination of high resolution in situ X-ray photoemission and X-ray absorption spectroscopies offer a powerful approach to monitor the deoxygenation process and comprehensively evaluate the electronic structure of Graphene Oxide thin films at different stages of the thermal reduction process. It is established that the edge plane carboxyl groups are highly unstable, whereas carbonyl groups are more difficult to remove. The results consistently support the formation of phenol groups through reaction of basal plane epoxide groups with adjacent hydroxyl groups at moderate degrees of thermal activation (~400 °C). The phenol groups are predominant over carbonyl groups and survive even at a temperature of 1000 °C. For the first time a drastic increase in the density of states (DOS) near the Fermi level at 600 °C is observed, suggesting a progressive restoration of aromatic structure in the thermally reduced graphene oxide.

**KEYWORDS.** Graphene Oxide, Thermal Deoxygenation, XPS, NEXAFS, Valence band, Oxygen-functional Groups and Atomic Structure, carbonyl, phenol.

## INTRODUCTION

Since the discovery of Graphene,[1,2,3] immense efforts have been focused on Graphene Oxide (GO),[4,5] since it is the most promising precursor for obtaining large quantities of this unique material and because GO is a useful material on its own right. Graphene Oxide can be visualized as individual sheets of graphene decorated with oxygen functional groups on both basal planes and edges, which has been prepared by oxidative exfoliation of graphite. The presence of oxygen makes GO amenable to chemical functionalization, nevertheless it disrupts the extended $sp^2$ network of the graphene hexagonal lattice. To convert GO back to graphene, the chemical/thermal reduction of GO is so far the most attractive procedure because of its simplicity, reliability, high yield and low cost.[6-14] Chemical treatment, especially exposure to hydrazine, is the most widely used route to reduce GO in solution.[7,9-12] However the chemically reduced GO (CRGO) suffers from a relatively low C/O atomic content,[12] with a considerable amount of residual O-moieties.[7,12] Also, hydrazine-treatment leads to the formation of nitrogen-functional groups,[7,12] along with the inherent toxicity of hydrazine. Improvements have been accomplished by either post-heating at low temperatures (200-500 °C),[8,10,11] or by replacing hydrazine by alternative less toxic reducing agents, e.g. sodium borohydride or alkaline solutions.[13,15] The use nontoxic and biocompatible reducing agents such as vitamin C (L-ascorbic acid),[16] green tea,[17] melatonin,[18] saccharides such as glucose, fructose and sucrose[19] or the use of environmentally friendly processes such as hydrothermal dehadration[20] have also been reported. In another approach,[21-23] the reduction of GO was achieved by a $TiO_2$ assisted photocatalytic method, in which the electrons photogenerated by UV–irradiated $TiO_2$ were injected into GO reducing the oxygen-containing functional groups. Such low temperature deoxygenating processes, although not completely effective in reducing GO, are amenable to electronic applications of graphene patterned onto glass or plastic substrates as well as to the synthesis of a wide range of functional hybrids with use in polymer composites, biosensors, energy storage and conversion technologies.[24] In comparison, thermal reduction at high temperatures (900–1000 °C), particularly in ultra high vacuum (UHV), is found to be highly efficient in producing graphene-like films with a significantly high C/O ratio,[12] with no introduction of any contaminants. However, such high-temperature processing is unlikely to be compatible with fabrication techniques used for most electronic applications. Practically the majority of the studies report the presence of various amounts of residual oxygen in reduced graphene oxide, with the electrical conductivity reaching values several orders of magnitude lower than that of mechanically exfoliated graphene.

To make further progress on optimizing and designing reduction processes, which is key to numerous applications, GO needs to be well characterized and its thermal deoxygenation needs to be well understood. Thermal reduction of GO has been shown to involve the removal of oxygen groups by formation of carboneous species ($CO_2$, CO) thus creating defects in the form of etch holes within the graphene basal plane.[10,11] There is however very little knowledge on how the oxygen containing functional groups of graphene oxide evolve during thermal reduction.[6,11,25] Theoretical and experimental studies have shown that formation of thermodynamically stable carbonyl and ether groups through transformation of the initial nearby hydroxyl and epoxy groups during thermal annealing, but little is known on the survival of the persistent residual oxygen groups.

The specific objectives of the present work were i) to elucidate the evolution of oxygen groups and probe their survival rate upon heat treatment; ii) to determine the nature of the residual oxygen containing functional groups that remain after reduction; iii) to clarify how the heat treatment affects the electronic structure of GO. To do that, we carried out high resolution in situ C 1s and O 1s core level and Valence band X-ray photoemission as well as X-ray absorption temperature dependent spectroscopic studies on GO. The analysis of these studies helped to develop a comprehensive view into the temperature evolution of electronic structure and surface chemistry of GO nanosheets. We found a predominance of phenol groups, which originate from the reaction of basal plane epoxide groups with adjacent hydroxyl groups, at moderate temperatures. It was established that these phenol groups survive even at temperatures of 1000 °C.

## EXPERIMENTAL

**Synthesis:** Highly oxidised Graphene Oxide (GO) was produced using a modified Hummers' process.[8] The starting material Graphite powder (product: 78391) with particle size ≤20 µm was purchased from Fluka and is denoted here as "pristine" graphite. All other chemical and reagents were purchased from Aldrich.

A mixture of 2.5 g of Graphite and 1.9 g of $NaNO_3$ was placed in a flask cooled in an ice bath. 85 mL of $H_2SO_4$ was added to the mixture and stirred until homogenized. Solution of 11.25 g of $KMnO_4$ in distilled water was gradually added to the solution while stirring. After 2 hours, the solution was removed from the ice bath, and further stirred for 5 days. Finally, brown-coloured viscous slurry was obtained. The slurry was added to 500 mL aqueous solution of 5 wt% $H_2SO_4$ over 1 hour while being continuously stirred. The mixture was stirred for a further 2 hours. Subsequently, 10 ml of $H_2O_2$ (30 wt% aqueous solution) was then added to the mixture and stirred for further 2 hours. This mixture was then left to settle overnight. The mixture was filtered and further purified by dispersing in 500 mL aqueous solution of 3 wt% $H_2SO_4$ and 0.5 wt% $H_2O_2$. After two days of precipitation, the supernatant solution was removed. This process was repeated five times. The solid product obtained after the rigorous cleaning process was rinsed using copious amounts of distilled water and dried in oven, as reported in literature.[8]. The resulting solid was dispersed in water by ultrasonication for 2 h to produce a GO aqueous dispersion. After one-day sedimentation, the thick flakes were removed and the supernatant was collected for further measurements.

**Characterization Techniques:** High-resolution transmission electron microscopy (HRTEM) analysis were carried out using JEOL 2100F, which has a point resolution of 0.19 nm. TEM samples were prepared on Holey carbon-coated Cu 300 mesh grids.

High resolution X-ray photoelectron spectroscopy (XPS) analysis was carried out using SCIENTA

ECSA 300 equipped with monochromatic Al Kα (h□ = 1486.6 eV) X-ray source at NCESS Daresbury Laboratory. Surface charging effects (due to insulating nature of as prepared GO) were compensated using a Scienta FG300 low energy electron flood gun at 4.0 eV. Step sizes of 1 eV and 0.05 eV were used for survey and high resolution spectra, respectively. Spectra were collected at room temperature and then at intervals of 200 °C up to 1000 °C. Heating for 2 minutes was done inside the chamber under UHV conditions of the order of $10^{-7}$ Torr using electron-beam heater, with electron beam impact on the back surface of the Si substrate. The samples were cooled before collecting the spectra. High-resolution in situ Valence band (VB) photoemission spectra were simultaneously collected at each temperature. For all spectroscopic studies, GO nanosheets were drop dried under infrared lamp to prepare thin films on Si substrates. Quantification was performed using the data analysis software (the ESCA300 data analysis software), associated with the SCIENTA ECSA 300 equipment, after performing a Shirley background correction. Calibration was carried by alignment of the spectra with reference to the C 1s line at 284.5 ± 0.2 eV associated with graphitic carbon. Binding energies were calibrated by the position of the Fermi cut-off of a gold foil for valence band data, and by the position of the Au $4f_{7/2}$ line (84.0 eV) in the case of core level data.

High resolution in situ Near-edge X-ray absorption fine structure (NEXAFS) spectroscopy was performed at Synchrotron Radiation Source (SRS), Daresbury Laboratory. Measurements were carried out at station 5U.1. Spectra at C K-edge and O K-edge were recorded in a total energy yield (TEY) mode at room temperature, 400 °C, 600 °C and 800 °C. All recorded spectra were normalized to the signal obtained from a gold covered grid.[26]

The initial GO sample was characterized by X-ray diffraction (XRD) and Raman spectroscopy, before and after the in situ temperature dependent XPS studies. XRD data were collected using a Philips 1050/81, for a step size of 0.02 and dwell time of 1 deg/min at standard potential and current settings of 40 kV and 20 mA, respectively, employing a monochromatic Cu Kα radiation source (λ = 1.54 Å). Raman spectroscopy was performed using Argon laser (λ = 514.78 nm) at an ISA Lab-Raman system.

Thermo-gravimetric analysis (TGA) was performed using a SDT Q600 V8.3 Build 101 system at a ramp rate of 1 °C/min up to 1000 °C in Nitrogen flow of 100 mL/min.

## RESULTS AND DISCUSSION

### Structural Characterization

#### *High Resolution Transmission Electron Microscopy (HRTEM) and Evidence of Graphitic C-backbone of GO*

HRTEM studies revealed the microscopic characteristics of as-prepared GO nanosheets, consisting of 2-4 layers (see Figure S1, Supporting Information), with limited sizes ranging from a few hundred nm to a couple of μm and a roughened surface due to the partial amorphous nature of the sample derived from the harsh oxidation steps involved in Hummers' method.[5,8] Figure 1a exhibits a typical TEM image of the triple-layered GO with the cross-sectional profile (bottom inset of Figure 1a). Despite the presence of such significant amount of O-species the long range orientational order is maintained. This is clear from the selected area electron diffraction (SAED) patterns (top inset of Figure 1a), where triple layers exhibited three sets of diffraction points. The occurrence of these misoriented hexagonal patterns implies an incommensurate stacking of the GO sheets. This is not surprising as the functional groups protruding from the GO planes are expected to decouple the interactions between the carbon backbones of neighbouring layers.

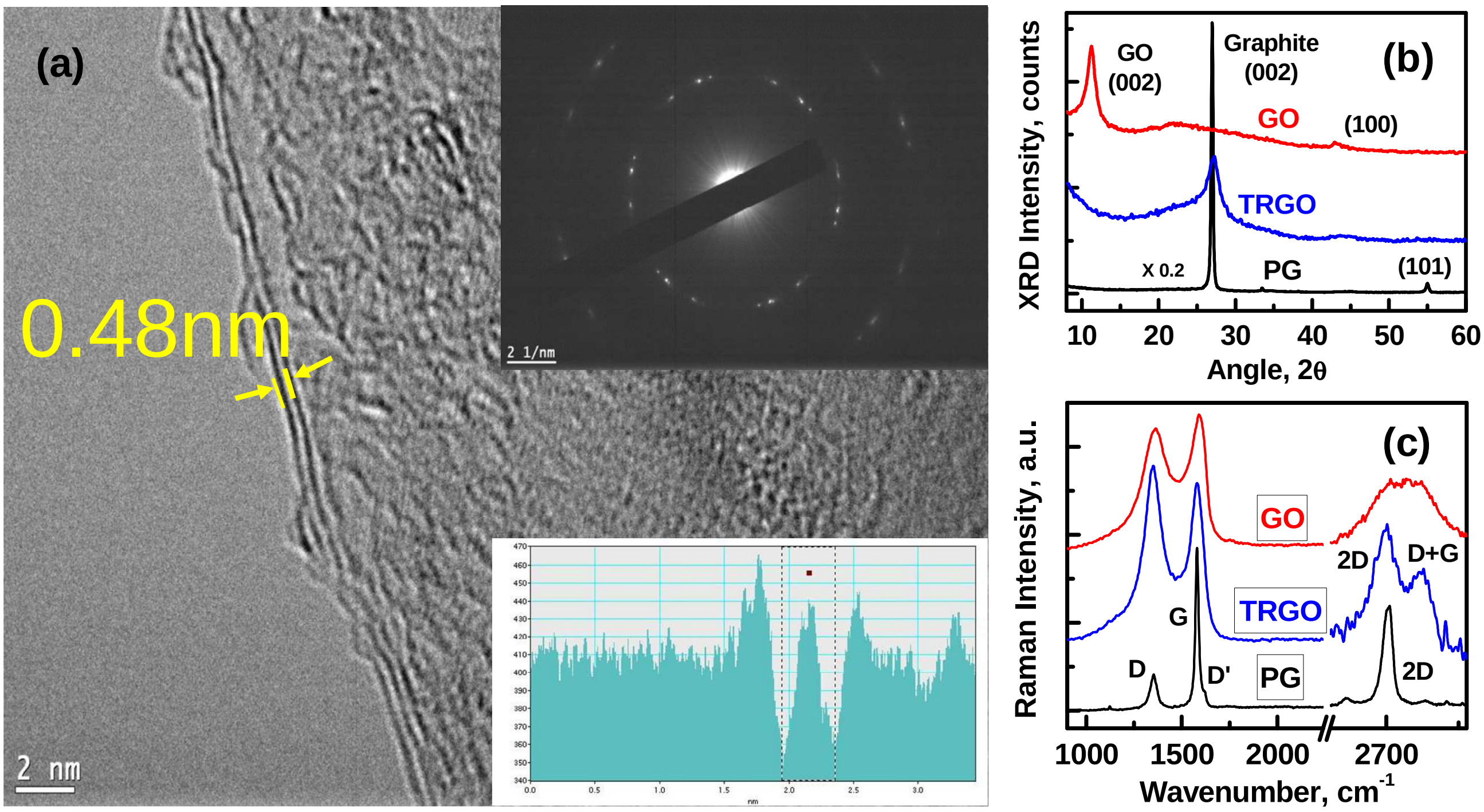

**Figure 1.** (a) TEM image of the triple-layered Graphene Oxide (GO) with the corresponding cross-sectional profile (bottom inset), and the SAED pattern (top inset). The arrows indicate a sheet separation of 0.48 nm. (b) XRD spectra and (c) Micro-Raman spectra of as-prepared GO and thermally reduced GO (TRGO) (vacuum-heat treated at 1000 °C), compared with that of pristine graphite (PG). The right panel of Figure 1(c) shows the magnified the 2D band region. The spectra were shifted in y-scale for clarity.

### *X-ray Diffraction (XRD)*

As evidenced from the XRD spectra (Figure 1b), the starting pristine graphite, PG, exhibits atomically flat pristine graphene sheets with a well-known van der Waals thickness of ~0.337 nm,[7,9,27] estimated by using Bragg's equation for the (002) peak located at ~26.4°. The (002) diffraction peak is found to be shifted for GO to ~11.3° indicating higher interlayer spacing (Δss = 0.782 nm). The highly oxidized GO sheets are expected to be 'thicker',[6,7,28] due to intercalated water molecules trapped between adjacent graphene oxide sheets,[29] with the Δss values been reported to vary from ~0.6 nm for ‘‘dry’’ GO to ~1.2 nm for “hydrated” samples.[7] The difference on the interlayer distance estimated by TEM (0.48A) and XRD (0.78 A) techniques can be explained bearing in mind the sample preparation and the environment in which the measurements were carried out. XRD measurements were performed on the samples comprised of thin films of GO in air, whereas TEM measurements were carried out in vacuum, on samples produced from dilute dispersion of GO on a TEM grid. The (002) XRD peak for GO shows considerably larger full width half maximum (FWHM), compared to PG. Upon thermal reduction (at 1000 °C in vacuum), the TRGO sample (originated from the GO) exhibits a structure closer to pristine graphite PG as revealed by the shifting of (002) peak back to 26.4° (Δss = 0.337 nm), even though its FWHM still remained larger than that of PG (Figure 1b), implicating the presence of strains/defects.

***Raman Spectroscopy***

Raman spectroscopy has played an important role for characterizing graphitic materials since it is able to provide information on crystalline size, the degree of hybridization, crystal disorder, the extent of chemical modification, and distinguish single layer graphene or nanotubes from multilayer ones.[30-38] The micro-Raman spectra (Figure 1c) of all the samples exhibited three main characteristics peaks: the G mode, a doubly degenerate (TO and LO) phonon mode ($E_{2g}$ symmetry) at the Brillouin zone center observed at ~1575 $cm^{-1}$, originating from in-plane vibration of $sp^2$ carbon atoms; the D mode arising from the doubly resonant disorder-induced mode (~1350 $cm^{-1}$) and the symmetry-allowed 2D overtone mode (~2700 $cm^{-1}$).[31-34] The GO sample shows a prominent D peak with intensity comparable to G peak, in sharp contrast to the smaller D peak of PG, indicative of significant structural disorder due to the O-incorporation. Additionally, the D band, attributed to in-plane A1g (LA) zone-edge mode, is innately Raman-active at the graphitic edges.[31-33,35] For small graphene sheets with limited sizes, like the GO nanosheets synthesized by the harsh Hummers’ method, the D band is expected to develop dramatically. Consequently, the sharp increase in $I_D/I_G$ ratio (from ~0.26 for PG to 0.93 for GO) indicates a decrease in the in-plane crystal or domain size[30] from ~17 nm (PG) to ~4.7 nm (GO). The G peak of GO is shifted to higher wavenumbers (~ 15 $cm^{-1}$) and broadens significantly with respect to that of graphite. Similar upward shifting of the G band has been observed in heavily oxidised carbon nanotubes[39] and was related to the emergence of a new Raman active band (D′ mode, ~1620 $cm^{-1}$) overlapped with the G band.[30] The D′ band, usually inactive, becomes Raman-active due to phonon confinement caused by defects.[40-42] Beside the influence of D′ band, Kudin et al.[35] have also considered the contributions from the isolated double bonds which yield Raman bands at little higher frequencies than the G band, for heavily oxidized GO. The vacuum-heat treatment (at 1000 °C) tends to shift back the Raman peaks (~ 12 $cm^{-1}$ red-shift of G peak with respect to GO), closer to the positions recorded for PG, indicative of the tendency to recover the hexagonal carbon network. The 2D band at ~2700 $cm^{-1}$ , which originates from a two phonon double resonance Raman process and is indicative of crystalline graphitic materials, exhibited the most interesting changes. Generally, the position and shape of the 2D peak are highly sensitive to the number of graphene layers, and has been utilised to distinguish the single-layer from few-layer graphene.[31-34] For the graphene samples prepared by micromechanical exfoliation of graphite[31,33] or by chemical vapour deposition[34] on thin metal (e.g. nickel) films, the single sharp 2D peak of monolayer graphene has observed to become wider and asymmetric with an upshift in peak position, with the increase in layer number. However in our study it is not possible to determine the number of layer in the thermally reduced GO, since the GO nanosheets were drop dried on a Si substrate to prepare a thin film. Hence the Raman spectra is the resultant signal of several stacked nanosheets, each one consisting of a few layers (2-4, as observed from TEM). The steep decrease in intensity and broadening of the 2D peak for GO compared with those of PG are mainly attributed to the steric effects of oxygen moieties on the stacked layers as well as to the partial amorphization and reduction in $sp^2$ domains.[31,38] Interestingly, the TRGO exhibits two distinguishable peaks close to 2700 and 2950 $cm^{-1}$ corresponding to the 2D band and to a (D+G) combination mode induced by disorder.[32] Such remarkable features, namely emergence of a sharp 2D band and of a lower intensity distinguishable D+G peak, vividly support the presence of smaller disorder in TRGO when compared with GO.

Interestingly, after annealing a slight increase in the D peak intensity is observed with the $I_D/I_G$ ratio increased to ~1.1, indicating a decrease in the size of $sp^2$ domains upon thermal reduction (~4 nm). Naturally, a decrease in this ratio would be expected upon annealing since the disorder associated with the oxygen-defects diminishes. Our experimental observations suggest that for the TRGO sample (heated at ~ 1000 C) even though the $sp^2$ sites are partially restored, the forced removal of oxygen at such high temperature leads to the creation of strains and/or topological defects on the C structure,[6,27] and hence the isolation of the $sp^2$ clusters forming smaller and dispersed $sp^2$ domains. Similar behavior i.e. slight increment or no change in $I_D/I_G$ ratio has been observed in a number of studies[7,43] involving post heating hydrazine reduction. Recent studies[36,41] have shown that the intensity ratio $I_{2D}/I_{D+G}$ may be a powerful indicator for the aromatic C-structural order of the graphitic materials, since the 2D band is sensitive to the aromatic C-structure, while the combination mode of (D+G) is lattice disorder induced band for crystalline graphitic materials.[37] Well-resolved 2D and (D+G) bands in TRGO sample and its higher $I_{2D}/I_{D+G}$ ratio (1.3 times higher than GO) indicate the restoration of aromatic C-structure upon thermal reduction of GO.[36,37]

## High Resolution in situ X-ray Photoelectron Spectroscopy (XPS) Analysis

### *Identification of Oxygen Moieties on GO structure*

Immediate observations from wide energy scan spectrum (WESS) of as prepared GO showed a clear shift in the XPS bands towards higher binding energy (BE) reflecting a significant surface charging effect due to the electrically insulating nature of GO (see Figure S2a, Supporting Information). After annealing, the oxygen content in TRGO becomes less than 2.9 at%, which is close to the value in graphite powder (3.4 at%), see Table S1 (Supporting Information). High-resolution C1s spectrum exhibited well-defined double peak formations, which is a signature of extreme oxidization in GO (at RT, Figure 2a). The assignment of C1s and O1s components were based on theoretical predictions of core level shifts and on reported spectra containing the particular oxygen functional groups. The XPS peaks were fitted to Voigt functions having 80% Gaussian and 20% Lorentzian character, after performing a Shirley background subtraction. In the fitting procedure, the full width at half-maximum (FWHM) values were fixed at a maximum limit of 1 eV for all the peaks. The $sp^2$ peak of the C1s envelope centred at 284.5eV had a FWHM of 1.0 ± 0.2 eV. It was found that the bands appearing at the higher energy region tended to be much broader (FWHM~1.7 eV) than the $sp^2$ component. In particular, the FWHM of the components at the tail of the C1s envelopes tended to be much wider than 2eV. In addition to the $sp^2$ graphite component at 284.49 eV, we found four broad components to account for the overlapping C1s features. The component at 285.86 eV is assigned to C atoms directly bonded to oxygen in hydroxyl configurations (shifts of 1–1.5 eV to higher BE). The component at 286.55 eV is attributed to epoxide group (C-O-C), and the smaller components at 287.54 and 288.94 eV are related to carbonyl (>C=O) and carboxyl groups (COOH or HO-C=O). The assignments are in agreement with the literature,[7,10-12,14,27,43-45] even though there is considerable vagueness and subjectiveness. One important issue is related to the presence of carbonyl >C=O groups: the basic model of GO electronic structure by Lerf–Klinowski et al.[29] has not accounted for any >C=O moiety, while De´ka´ny model[46] and later Ajayan et al.[15] identify its contribution. On the contrary, reports by Jeong et al.[27,47] claim the absence of any experimental evidence of >C=O group, accepting Lerf–Klinowski model,[29] which assumed the doubly bonded oxygen C=O species exist only as part of the COOH groups at the edge sites of GO sheets. In a number of reports on GO, the deconvolution of C1s spectra has been performed using four components, namely $sp^2$, C-OH, C-O-C and COOH, ignoring the presence of >C=O groups.[9,27,29,47] Whereas, other reports[7,10,11,14] consider only one peak for singly bonded oxygen C ─ O groups, performing 4-peak-deconvolution for $sp^2$, C ─ O, >C=O and COOH.

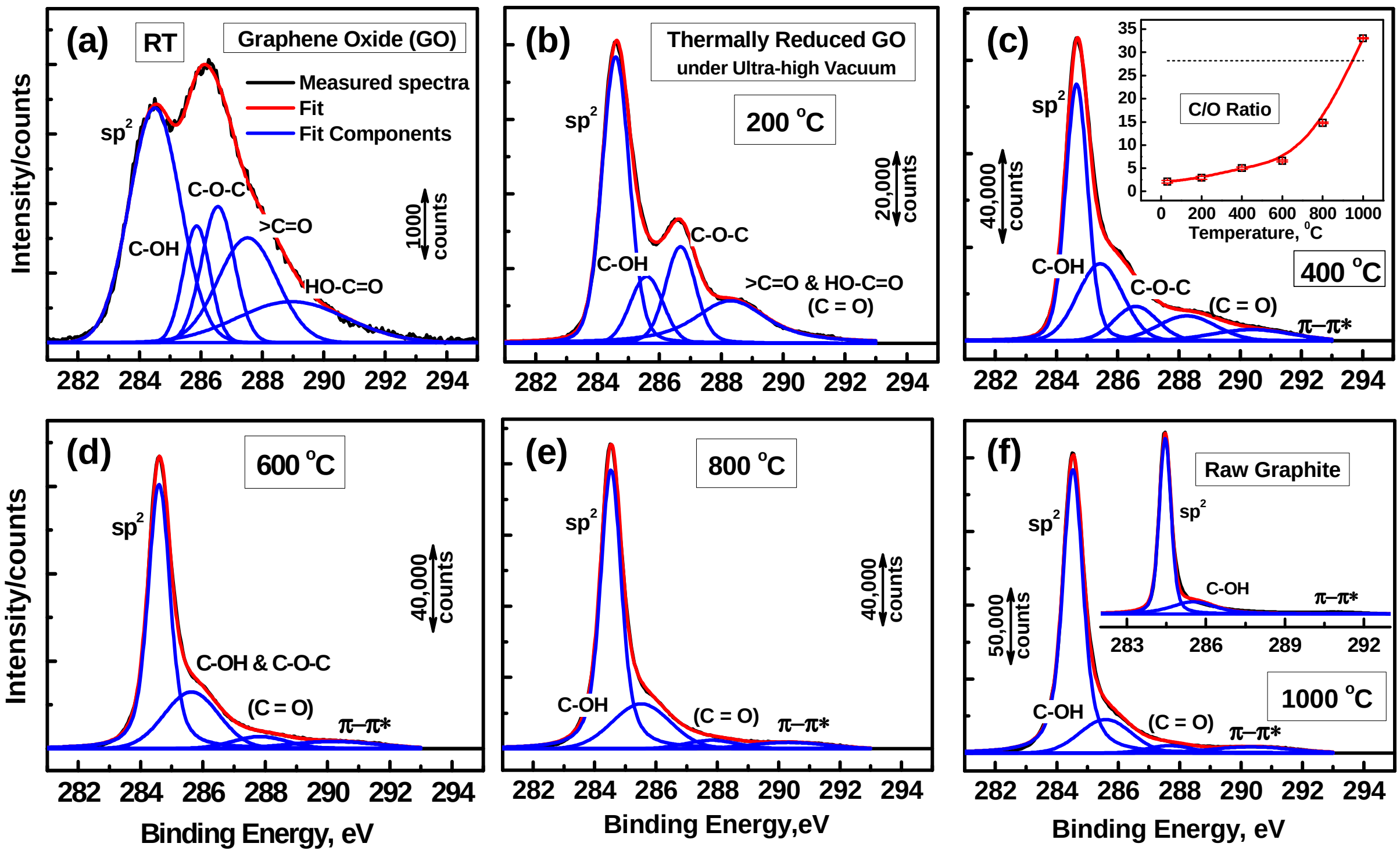

**Figure 2.** High resolution C1s XPS spectra: deconvoluted peaks with increasing reduction temperature (Tr). Inset of Figure 2c: C/O ratio as a function of Tr (Error bars represent the standard deviation estimated from six sets of data). Inset of Figure 2f: C1s spectra for starting/precursor graphite (PG).

The symbols specify the following groups:

C-OH ⇒ Hydroxyl, C-O-C ⇒ Epoxide, >C=O ⇒ Carbonyl, HO-C=O ⇒ Carboxyl.

C ═ O ⇒ Oxygen doubly bonded to Carbon, C ─ O ⇒ Singly bonded Oxygen.

π→π* ⇒ Shake-up satellite peak, $sp^2$ ⇒ C to C bond in aromatic rings.

The starting graphite (PG) shows 2 main peaks, namely $sp^2$ (284.47 eV) and C-OH (285.53 eV) (inset of Figure 2f). The presence of a weak C ─ O peak (~285.5 eV) in graphite, associated with atmospheric oxidation, has previously been observed by Hontoria-Lucas et al.[43] Furthermore, Barinov et al.[44] assigned the same peak at ~285.6 eV to C ─ O single bond, calculating its close chemical shift of ~1-1.5 eV to higher BE relative to $sp^2$ peak.

Here we have judiciously assigned the peak at ~285.5 eV to hydroxyl/phenolic group, and its neighbouring peak at ~286.5 eV to epoxy group, since it should have a larger BE compared to hydroxyl groups.[12,27,45] The C ═ O double bond emission occurs at even higher BE range and arises from >C=O (~287.5 eV) followed by COOH (~289 eV).[12,45]

### *Contribution of Oxygen Moieties on GO structure*

In pristine GO, the C/O atomic ratio, calculated by dividing the area under C1s peak with that of O1s peak-area and multiplied by the ratio of photo-ionisation cross sections, was found to be only 2.08 (inset of Figure 2c), with a C contribution of ~67.5% (Table S1). Another important parameter that can be used to characterize the degree of oxidation in GO is the $sp^2$ carbon fraction, which was estimated by dividing the area under $sp^2$ peak with that of C1s peak-area. We found, the $sp^2$ fraction of GO is only 40% (Figure 3a). Carbon atoms connected with hydroxyl and epoxy groups are $sp^3$ hybridized. In the basal plane carbon atoms bonded with C-O-C (epoxides) prevail over (1.5 times higher) the hydroxyl C-OH groups (Figure 3), in agreement with molecular dynamic simulations, according to which the ratio of epoxides to hydroxyls increases with increasing the oxidation.[10,11,28] The contribution of >C=O and COOH is found to be more substantial compared to C-OH and C-O-C ([>C=O + COOH] / [C-OH + C-O-C] = 1.4). This can be understood bearing in mind that at harsh oxidisation conditions, such as those encountered in Hummers' process, the oxidisation of C ─ O single bonded groups to C ═ O double bonded species is energetically favorable,[15,27,46] Interestingly, maximum contribution is found from >C=O groups. Their profusion in GO supports their definite existence on GO as predicted by the De´ka´ny model,[46] which identified >C=O contributions in the form of ketones/quinines, updating the Scholz-Boehm[48] and Hontoria-Lucas[43] models. More recently, Ajayan et al.[15] suggested the possible generation of ester carbonyls through the reaction of tertiary alcohols with nearby carboxylic acids, at high degrees of oxidation.

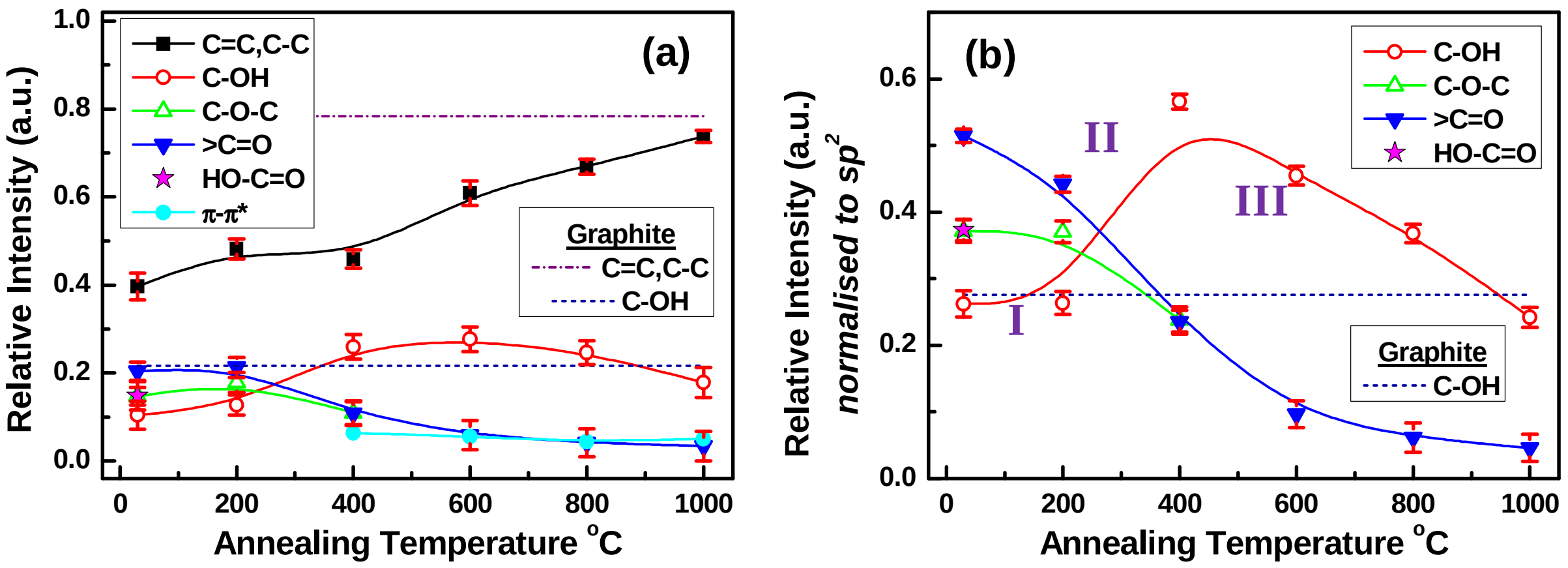

**Figure 3.** Temperature-dependence of (a) the relative contribution of C1s peak components estimated by dividing the area under each component by whole C1s peak-area, and (b) the normalised intensity of O-groups relative to $sp^2$ intensity. The lines shown are guides to the eye only. Dotted lines represent the values corresponding to C1s peak components, observed for graphite. Error bars represent the standard deviation estimated from six sets of data.

### *Thermal Evolution of GO*

***Restoration of Aromatic C-structure*:** Upon heating under UHV, the C1s spectrum exhibits a transformation from a double peak at room temperature to a single sharp peak (~284.5eV) at 1000 °C, resembling the C1s peak of PG and been indicative of a trend to restore the $sp^2$ bonding graphene character (see Figure S2b). A clear shift of peak-maxima back to lower BE with increasing temperature signifies the transformation of electrically insulated GO to the conducting nature of graphite. The evolution of the C/O atomic ratio (inset of Figure 2c and Table S1) reveals an increase in C-content as reduction temperature (Tr) increases, and an associated decrease of the O groups. A maximum C-content of ~97% can be achieved upon heating at 1000 °C, with only ~3% of remnant O contribution (C/O ratio ~ 33.02), alike PG (Table S1). The (π→π*) shake-up satellite peak, observed for PG (inset of Figure 2f) around ~290.5 eV, appears upon heating at high temperatures ≥ 400 °C (Figure 2c-f). This indicates that the delocalized π conjugation, a characteristic of aromatic C structure, is to some extent restored in TRGO samples.[10,11,15]

***Thermal Stability of Hydroxyl Groups*:** Figures 3a and 3b present (i) the relative contribution of the carbon bonds in GO and (ii) the normalized intensity of O-related peaks (relative to $sp^2$ peak-intensity) as a function of annealing temperature in UHV respectively. These figures serve as useful guides for monitoring the evolution of the functional groups and provide insight into the mechanism of thermal reduction process. At room temperature the determination of separate >C=O and COOH contributions was possible, however for temperatures higher than 200 °C, we denote their combined involvement as C ═ O (Oxygen doubly bonded to Carbon), since the deconvolution into two separate peaks was not feasible. The C ─ O component (singly bonded Oxygen, C-OH and C-O-C) remains almost unchanged (Figure 3b, region I) up to 200 °C, most possibly due to insufficient temperature and/or partial contribution from the transformation of C ═ O to C ─ O. Above 200 °C, C ═ O continues to be reduced, and around 800 °C, it almost saturates with a minimal contribution.

Useful information can be obtained from the dispersion of the peak positions as a function of temperature. It is worth pointing out that at 200 °C, the C ═ O peak is centered at ~288.3 eV, between the >C=O (~287.5 eV) and COOH (288.9 eV) components (Figure 2a and Figure 4a). However at high temperatures it shifts to BE representative of >C=O. Therefore it is reasonable to postulate that at low reduction temperatures (Tr < 400 °C) the doubly-bonded C ═ O component has contributions from both >C=O and COOH groups, however at higher temperatures (Tr > 600 °C) is dominated by >C=O groups.

In Figure 2b one can note that at Tr >200 °C the C-O-C is reduced fast, and at Tr ≥ 400 °C, it is hard to be identified (Figure 2d-f). Interestingly, over the temperature range 200-400 °C the C-OH increases rapidly (Figure 3b, region II), followed by a reduction at Tr > 400 °C (Figure 3b, region III). C-OH persists even at 1000 °C, at a level similar to that observed in the original graphite material PG (blue dotted line, Figure 3b). Our observations are consistent with other reports, where the C-OH contribution survives in TRGO showing that annealing at 1000 °C is not adequate to completely remove the oxygen,[10-12,14] even though possessing higher C/O ratio (~ 33.02). Complete thermal pyrolysis of C-OH is thermodynamically difficult,[28] mainly due to its intercalating position into the interlayer galleries between intact conjugated domains.[46]

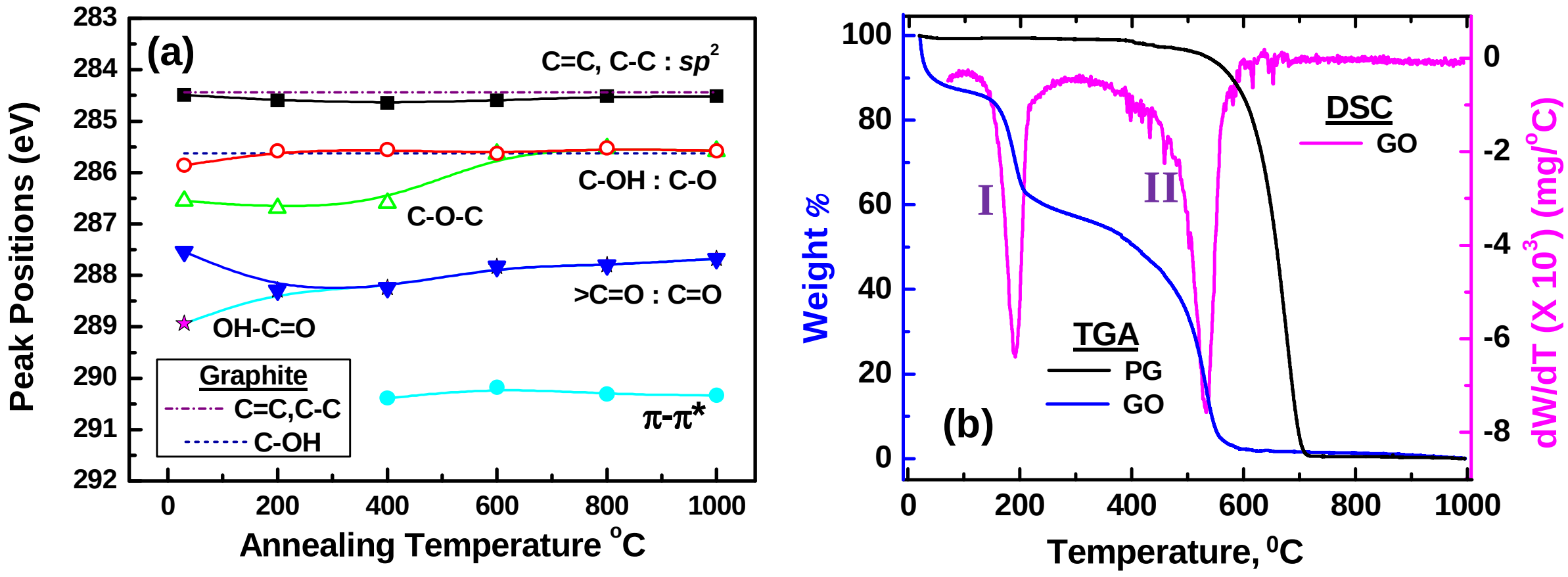

**Figure 4.** (a) Temperature-dependence of the peak-position of C1s peak components; dotted lines represent the values corresponding to PG. The lines shown are guides to the eye only. (b) TGA results for GO and PG together with corresponding mass loss rate (dW/dT) for GO.

### *Thermo-Gravimetric Analysis (TGA) Studies*

Interestingly, the anomalous trend of C-OH (Figure 3b) clearly implicates a double transition zone on the reduction path of GO. The phenomenon is vividly illustrated in TGA results (Figure 4b). The GO starts to lose mass upon heating even below 100 °C, which is associated with elimination of loosely bound or adsorbed water and gas molecules. First, major mass loss can be observed along with an exothermic signal of mass loss rate (dW/dT, I, Figure 4b) around 200 °C, yielding CO, $CO_2$ and steam as by-products of the reduction process. Second, the largest mass loss, and corresponding exothermic dW/dT signal, starts at Tr > 300 °C and continues till 600 °C, (II, Figure 4b). The higher curvature in TGA curve and the asymmetric nature of dW/dT signal, at 300-500 °C, implies the presence of two antagonistic activities: either (i) release of by-products and subsequent trapping and/or (ii) loss of doubly-bonded C ═ O component and simultaneous feeding of C ─ O singly-bonded oxygen groups. In the same phase, C-OH also shows an increase (Figure 3b, region II) while C ═ O species decrease. Above 500 °C, higher slope in TGA & sharp-rise in dW/dT signal indicate a rapid decomposition of O-species, as observed in Figure 3b (region III).

***Anomalous Trend of Hydroxyl Groups and Formation of Phenol***

Such anomalous trend of temperature evolution of C-OH has never been reported before, though Lerf et al.[29] have already predicted the formation of phenol (or aromatic diol) groups during deoxygenation, even at 100 °C, because of the close proximity of C-O-C and C-OH on the basal plane. Presence of enolic OH species is also considered by De´ka´ny et al.[46] in order to interpret the planar acidity of GO. Hence, it can be envisaged that initial rise of C-OH is mainly contributed by newly formed phenolic groups in expense of C-O-C, which show huge loss at Tr ≥ 400 °C. In addition, though C-O-C and C-OH become inseparable at Tr ≥ 600 °C, Figure 4a clearly shows that singly bonded C ─ O group retains its peak-position at that of C-OH, indicating the complete loss/conversion of C-O-C.

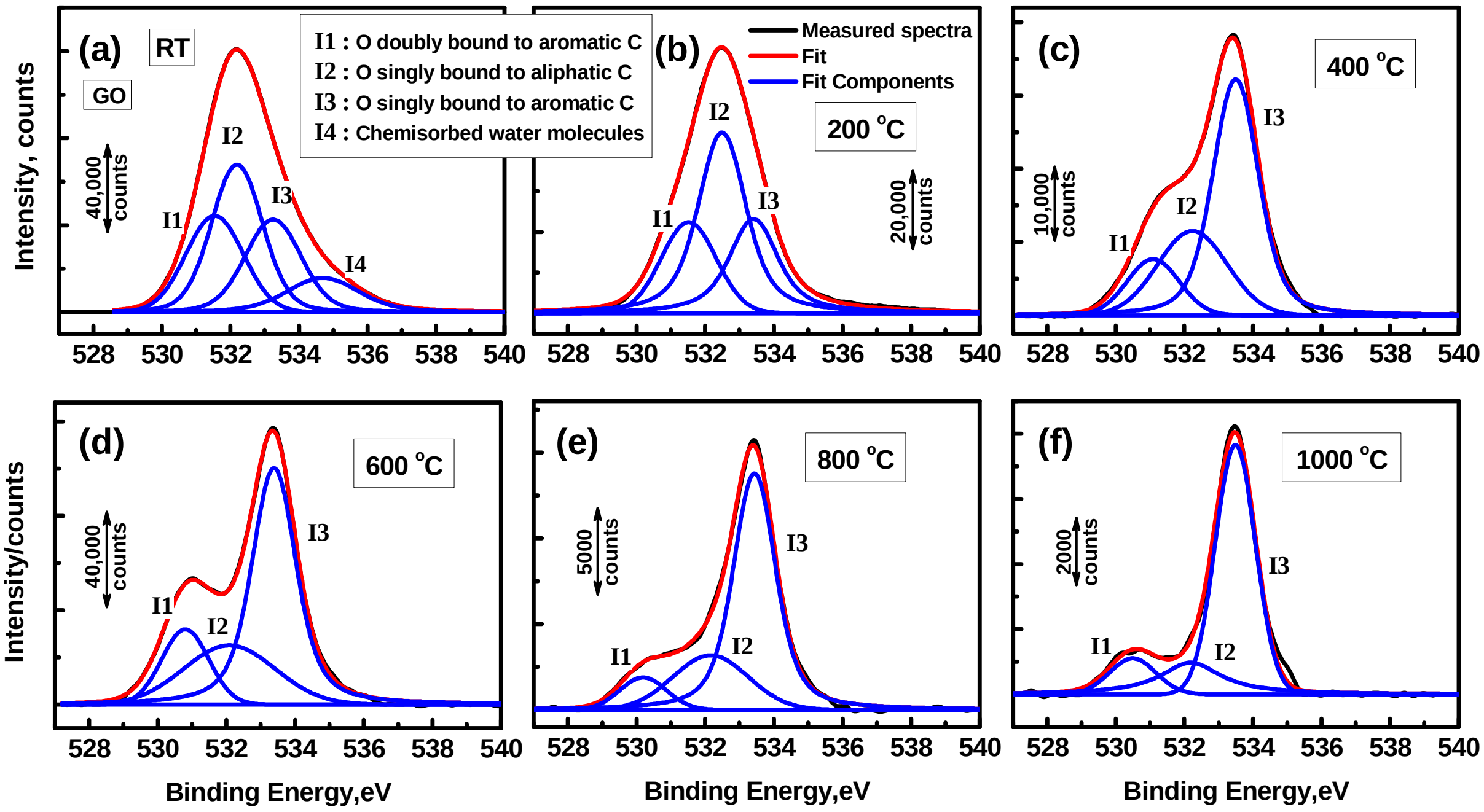

**Figure 5.** High resolution O1s XPS spectra deconvoluted peaks with increasing Tr.

Formation of phenolic groups is clearly evidenced from the high resolution O1s peaks (Figure 5). Deconvolution of O1s spectra (Figure 5a) produces 3 main peaks around 531.08, 532.03 and 533.43 eV assigned to C = O (Oxygen doubly bonded to aromatic Carbon-denoted as I1),[10,11,14,44] C − O (Oxygen singly bonded to aliphatic Carbon- denoted as I2),[6,43] and Phenolic (Oxygen singly bonded to aromatic Carbon- denoted as I3) [6,43] groups respectively. The pristine GO shows an additional peak at higher BE (I4 ~534.7 eV) corresponding to the chemisorbed/intercalated adsorbed water molecules.[14] Thermal treatment of GO clearly causes a shift of O1s spectra to the higher energy side and a simultaneous transformation of O1s spectra from a single peak to a double feature with the development of a prominent phenolic I3 peak (Figure 5). Although all the O-species decrease in integrated intensity with increasing Tr (Figure 6a), the relative intensity of phenol group (peak I3) shows a sharp rise around 400 °C (Figure 6b), as observed in C1s spectra (Figure 3), followed by a progressive increase relative to other species (Figure 6b). In addition, the relative intensity of peak I2, O singly bound to aliphatic C, exhibits an initial increase at 200 °C and a subsequent reduction. The initial increase is most possibly due to the internal conversion of C ═ O to C ─ O at low Tr, as described earlier. Undoubtedly, the evolution of O1s spectra provides further corroborating evidence on the existence of double transition zone during thermal reduction of GO.

The XPS analysis reveals the OH-moieties on basal plane are the most thermally stable species, in contrast to observations[7,9] made on chemical reduction of GO. Further corroborative evidence on the evolution of the basal and edge plane O groups, upon thermal reduction, is provided by NEXAFS studies.

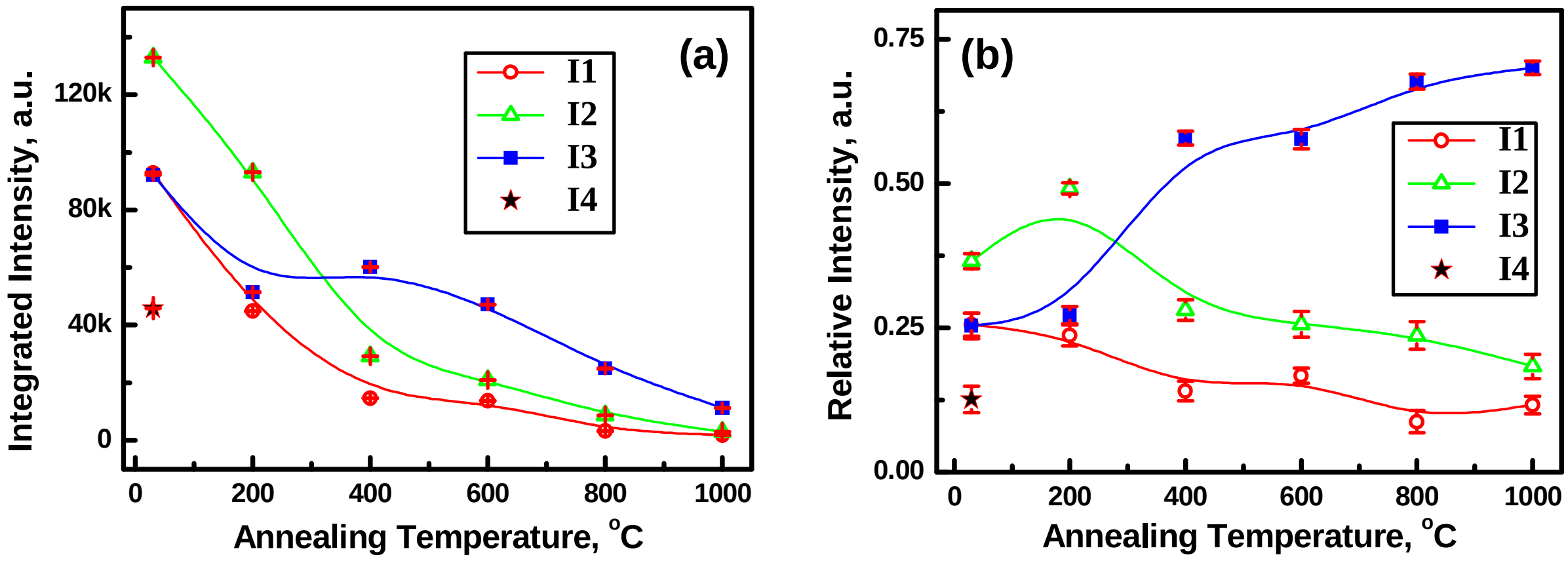

**Figure 6.** Temperature-dependence of (a) the integrated intensity of O1s peak components (I1-I4), and (b) the corresponding relative contribution estimated by dividing the area under each component by whole O1s peak-area. I1 denotes C ═ O (Oxygen doubly bonded to aromatic Carbon); I2 denotes C ─ O (Oxygen singly bonded to aliphatic Carbon), I3 denotes Phenolic (Oxygen singly bonded to aromatic Carbon) and I4 denotes chemisorbed/intercalated adsorbed water molecules groups. Error bars represent the standard deviation estimated from six sets of data. The lines shown are guides to the eye only.

## High Resolution in situ Synchrotron Near-Edge X-ray Absorption Fine Structure (NEXAFS) Spectroscopy

Soft X-ray absorption spectroscopy probes unoccupied electronic states and is another powerful tool for characterising graphitic materials, by providing information on the degree of bond hybridization in mixed $sp^2/sp^3$ bonded carbon, the specific bonding configurations of foreigner functional atoms and the degree of alignment of graphitic crystal structures. Fingerprints of the species surviving at each step of the thermal treatment were provided by NEXAFS. Here, NEXAFS was deliberately performed at 90° incidence of the linearly polarized X-rays. At normal incidence of the polarized X-ray beam the electric-field vector E lies within the graphene plane, and thus transitions to states of σ symmetry are more prominent than those to π symmetry.

The high resolution C K-edge NEXAFS spectrum (Figure 7a), of pristine GO, provides a clear presence of both unoccupied π* (1s→π*) and σ* (1s→σ*) states around 285.2 and 293.03 eV, respectively,[49] revealing that GO nanosheets, even though highly oxidized, still maintain the aromaticity of the original pristine material, PG. Although GO produces a plethora of O-related resonances, unfortunately, NEXAFS database in literature is not rich enough to deconvolute and assign all the peaks. Upon judicious review of the available literature we have assigned the peak-positions,[9,47,49-56] as shown in Figure 7a.

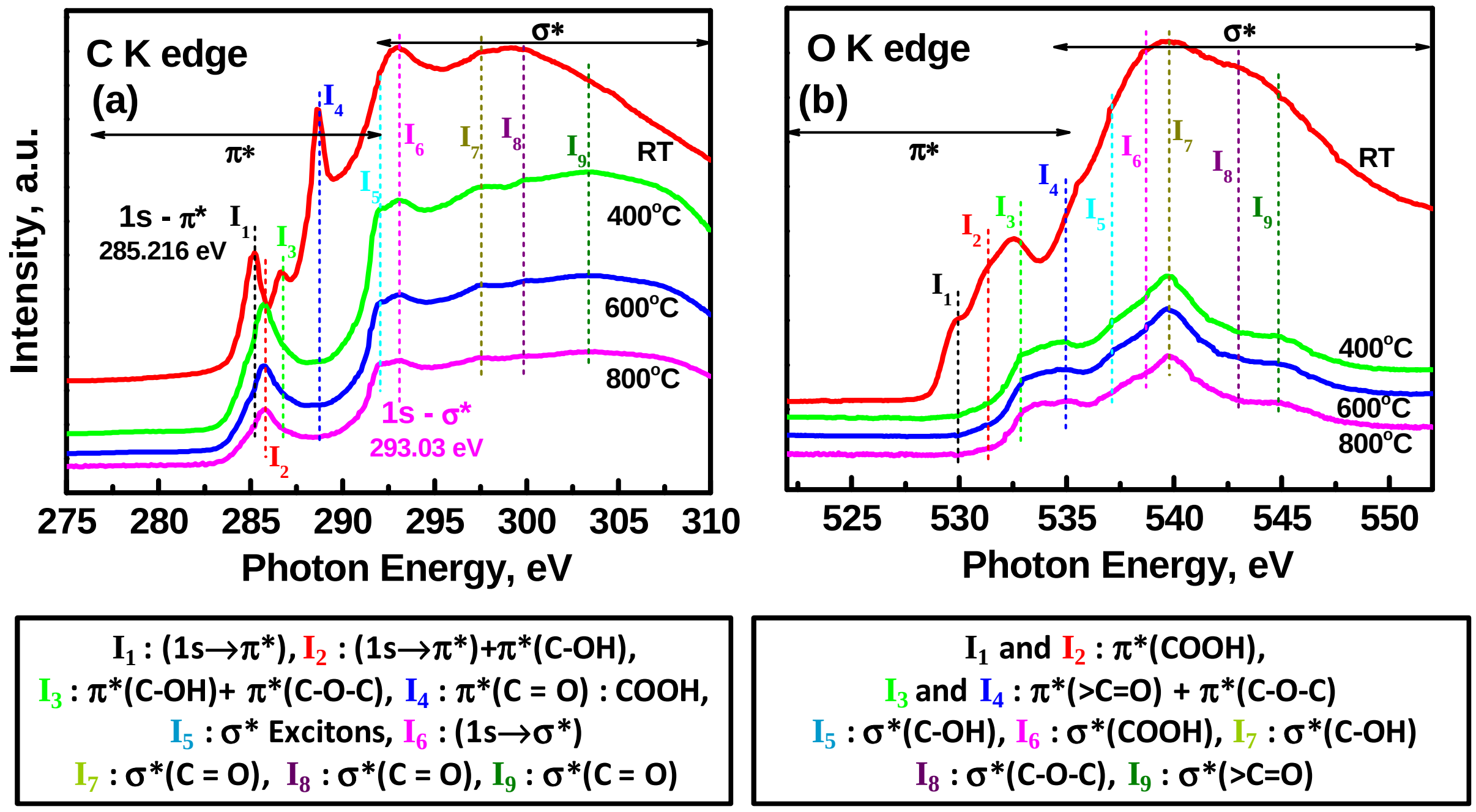

**Figure 7.** High Resolution (a) C K-edge and (b) O K-edge synchrotron NEXAFS spectra, recorded at different reduction temperatures. The spectra were shifted in y-scale for clarity.

Upon thermal treatment, a number of important changes can be identified in the spectra of TRGO. The π* resonance ($I_1$) clearly shifts to higher energies, moving towards the position of $I_2$, whereas its FWHM becomes broader due to presence of mainly C-OH moieties. The two resonances between the π* and σ* discernible at ~287 ($I_3$) and ~288.7 eV ($I_4$) disappear at 400 °C. The assignment of these intermediate peaks in the absorption spectra of graphene related structures is highly debatable. In particular, the presence of peak around 288 eV was originally observed in the NEXAFS spectra of HOPG[55] and few-layer graphene[56] and was attributed to the free electron like set of bands corresponding to electronic excitations lying between graphite layers (interlayer states). Others have provided evidence that this feature (~289 eV) originates from –COOH moieties present in single-wall carbon nanotubes,[53] carbon fibres[49] and GO.[9,47] Here, given the prominence of $I_4$ peak in heavily oxidised graphene films and its complete disappearance upon thermal annealing, we have assigned its origin to COOH consistent with XPS, where the presence of a sufficient density of carboxylic moieties in GO has been confirmed. The broad asymmetric peak $I_3$ originates mainly from π*(C-O-C) contributions and partly by π*(C-OH).[9,47]

The σ* region, which is strongly enhanced due to normal incidence of the polarized beam is dominated by doubly bonded C = O moieties (peaks $I_7$ to $I_9$).[53,54] Its temperature evolution reveals a huge loss of C = O groups, which agrees well with our earlier XPS discussion on thermal instability of C = O groups. At 400 °C, there is a curious splitting of the σ* peak (1s→σ*) into 2 peaks, with the simultaneous emergence of the σ* exciton peak ($I_5$).[49] The appearance of the distinct peak $I_5$ supports the progressive restoration of graphitic structure in GO, however its reduced/sub-expressed structure, compared to π* peak, indicates the existence of small graphene planes/domains,[49] as predicted in XRD and Raman studies.

Remarkably, after thermal reduction, transitions to states of σ symmetry are seen to decrease dramatically compared to transitions to states of π symmetry. This is visualized in Figure S3, which depicts the intensity ratio of π*/σ* peaks at the C K-edge, as a function of annealing temperature. This ratio is used to provide an estimate of the relative concentration of $sp^2$ domain configurations in a $sp^3$ matrix consisting of carbon atoms connected to oxygen groups.[51] High values of π*/σ* ratio around the temperature zone of 400–600 °C due to the substantial reduction of edge plane doubly bonded C ═ O moieties are consistent with the appearance of aromatic characteristic (π→π*) plasmon peak at the same Tr-range. At higher temperatures Tr > 400 °C, the change in π*/σ* is less dramatic due to slow reduction in both edge and basal plane O-moieties.

Further evidence on the significant loss of oxygenated functional groups upon thermal treatment at 400 °C is provided by the O K-edge NEXAFS spectra presented in Figure 7b, where the main peaks are assigned according to literature.[9,47,50-52] Two distinctive peaks $I_1$ and $I_2$ appearing at the low energy tail of O K-edge attributed to the π* state of COOH groups located at the GO edge sites,[9,47] disappear at 400 °C.

**High Resolution in situ Valence Band (VB) Spectroscopy**

Valence band spectroscopy is another powerful tool[13,57,58] to evaluate how the π conjugated system has been progressively restored after thermal treatment. Figure 8a shows the high-resolution VB spectra recorded at different temperatures. Based on photoemission spectroscopic data and theoretical band structure calculations for graphite, the region from 2 to 12 eV above Fermi level ($E_F$) represents characteristics of C2p electrons, the section 12-22 eV corresponds to C2s valence electrons, followed by the O2s region at higher BE.[44,59-62]

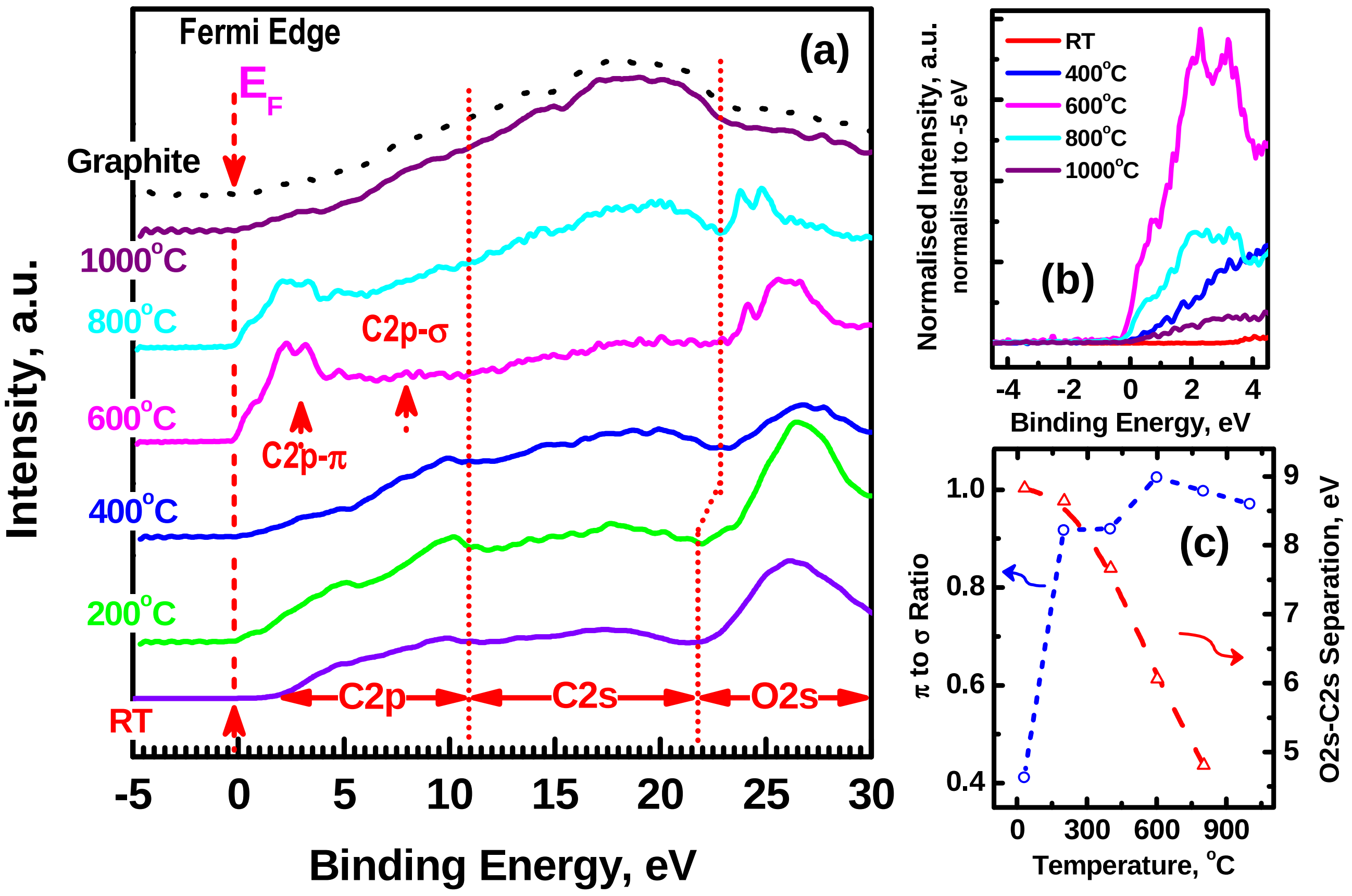

**Figure 8.** (a) High Resolution Valence Band (VB) spectra, recorded at different Tr. The spectra were shifted in y-scale for clarity. (b) Enlarged view of VB spectra at the vicinity of Fermi level ($E_F$). (c) Intensity ratios of C2p-π to C2p-σ (π to σ) and the separation between O2s and C2s peak-maxima, as a function of Tr. The lines shown in (c) are guides to the eye only.

A number of interesting features can be identified from the VB spectra. As prepared GO is dominated by O2s with a band centred at 26.3 eV[61] and its Fermi edge shifted to higher energies revealing the insulating nature of the material. The shoulder located at ~8 eV is associated with strong σ bonding states (C2p-σ) in graphitic like carbon, indicating the presence of a substantial $sp^2$ ordering.[44,59,60] Thermal reduction improves the contribution from C2p states and simultaneously reduces the O2s confirming the deoxygenation of O-moieties. In particular thermal treatment at 600 °C, induces the growth of a new sharp double feature with peaks at 2.3 and 3.25 eV associated with conjugated π bonds (C2p-π bands) of graphene.[44,59,60] An enrichment of the π-peak intensity in the valence band photoemission data implies the formation of sizable graphene domains with 3-fold coordination.[60] VB spectra at 1000 °C resemble those of the starting Graphite.

Interestingly, the π-derived density of states (DOS) at the vicinity of $E_F$ in the range of 0-2.0 eV (Figure 8b) rises with increasing temperature, with the steepest rise occurring at 600 °C. This rise in VB provides a strong indication for the existence of metallic character in the reduced graphene.[44,62] The simultaneous rise of DOS at $E_F$ and the enrichment of π-peak intensity in C2p region definitely reflect the progressive restoration of π conjugated aromatic C-structure by de-oxidation via thermal treatment. Curiously, further increase in temperature (> 600 °C) caused a reduction in the π-derived DOS indicating the presence of a high level of defects. The observation is consistent with Raman and XRD results, which confirmed that thermal treatment at 1000 °C is not adequate for the complete restoration of aromatic C-structure. Presumably the enforced removal of basal O-species produces strains in C=C bonds and topological wrinkles or hole-like defects on the atomic C structure.[6,27] Consistent with the XPS results the π/σ ratio obtained from C2p-π/C2p-σ bands shows an initial drastic increase followed by a slower increase (Figure 8c).

Another important observation is the progressive decrease in the O2s-C2s peak separation with thermal treatment illustrated in Figure 8c. The augmentation of O2s-C2s separation has been associated with the presence of intermediates between the C ═ O double bond groups and the C ─ O single bonded O to C, in progressively oxidized graphitic materials, following the order: C–O–C > C ═ O > C–OH.[61] Thus, the smallest separation, observed in highly reduced GO (~800 °C), would be attributed to the prevalence of phenolic/hydroxyls over other O-species.[61] The larger separation at low temperatures would indicate contribution from C-O-C and C ═ O moieties.[61]

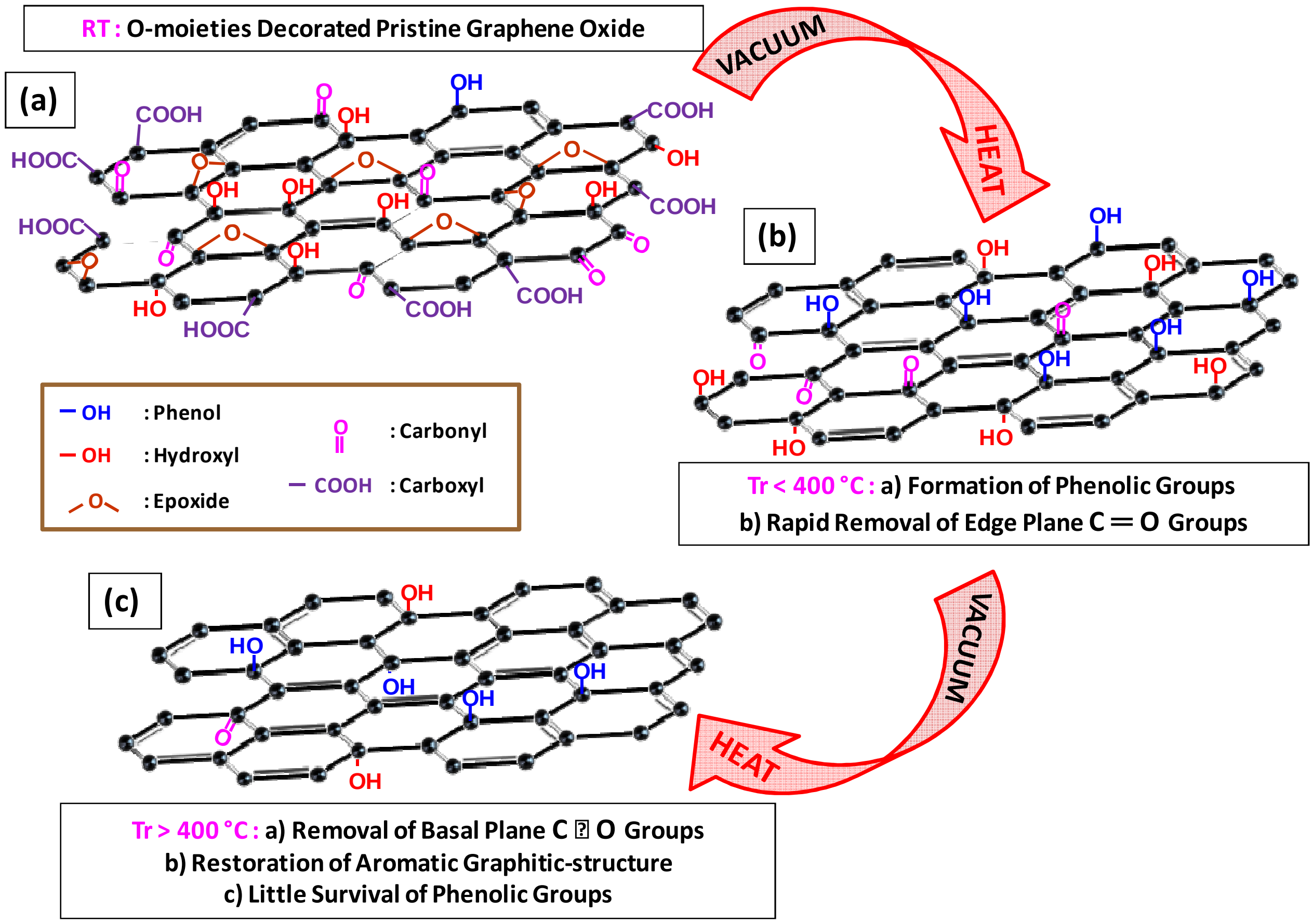

**Figure 9.** Schematic diagram of the temperature evolution of GO.

## CONCLUSIONS

Employing a combination of high resolution in situ temperature dependent spectroscopic techniques including XPS, VB and NEXAFS we have clarified a number of important issues concerning the evolution of the electronic structure of GO upon the heat treatment. (i) First it is established, that upon progressive thermal treatment, the edge-plane COOH groups become highly unstable, whereas >C=O are more difficult to be removed. (ii) The C-OH group in the phenolic form is the most thermally stable of all the oxygen species (Figure 9c). (iii) The thermal evolution of C-OH groups exhibits a well-defined transition temperature around 400 °C. At lower temperatures, there is an upward trend due to the formation of phenol groups whereas at Tr > 400 °C there is decreasing trend due to their decomposition (Figure 9b-c). (iv) Valence band spectra reveal a drastic increase in the DOS near $E_F$ at 600 °C, whereas further increase in temperature reduces the DOS to levels similar to those of graphite. The ability to achieve higher DOS at considerably lower temperatures compared to 1000 °C normally considered for partially restoring the aromaticity of thermally reduced Graphene Oxide holds significant advantages for nanoelectronics application.

**ACKNOWLEDGMENT.** The work was supported by The Leverhulme Trust (fellowship for Dr Ganguly: 1-212-R-0197), The Royal Academy of Engineering, EPSRC funded facility access to HRTEM at University of Oxford (EP/F01919X/1), NCESS XPS facility (EP/E025722/1) in Daresbury, and NCESS Synchrotron facility in Daresbury. Also the assistance from Mr. S. Ukleja at FSERTC (UU); and Dr. D. S. Law and Dr G. Beamson at NCESS is acknowledged. Supporting Information is available online at http://pubs.acs.org.

**REFERENCES.**

**SYNOPSIS TOC.**

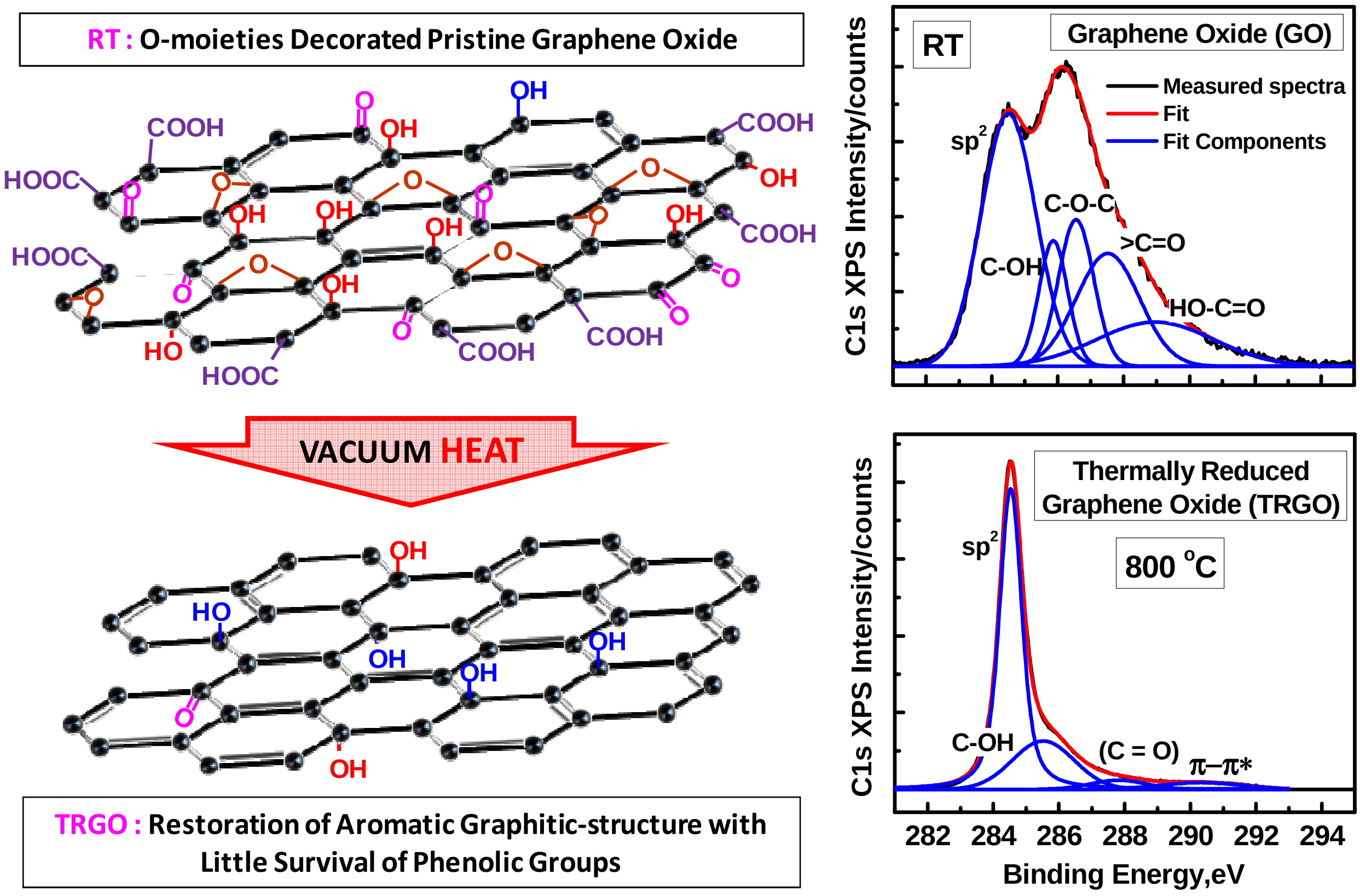

**Supporting Information**

Probing the Thermal Deoxygenation of Graphene Oxide using High Resolution *In Situ* X-Ray based Spectroscopies

*Abhijit Ganguly,[§] Surbhi Sharma,[§] Pagona Papakonstantinou*, and Jeremy Hamilton*

Nanotechnology and Advanced Materials Research Institute, NAMRI, University of Ulster, Jordanstown campus, BT37 0QB, United Kingdom. [§]These authors contributed equally.

*Corresponding author. E-mail: p.papakonstantinou@ulster.ac.uk

## Supplementary Figures

**Table S1.** High resolution XPS analyses: atomic percentage of carbon (C) and oxygen (O), estimated from area under C1s and O1s peak, respectively, as a function of temperature.

| | **RT** | **200°C** | **400°C** | **600°C** | **800°C** | **1000°C** | **PG** |
|---|---|---|---|---|---|---|---|
| **C** | **67.54** | **74.52** | **83.42** | **86.80** | **93.66** | **97.06** | **96.58** |
| **O** | **32.46** | **25.48** | **16.58** | **13.20** | **6.34** | **2.94** | **3.42** |

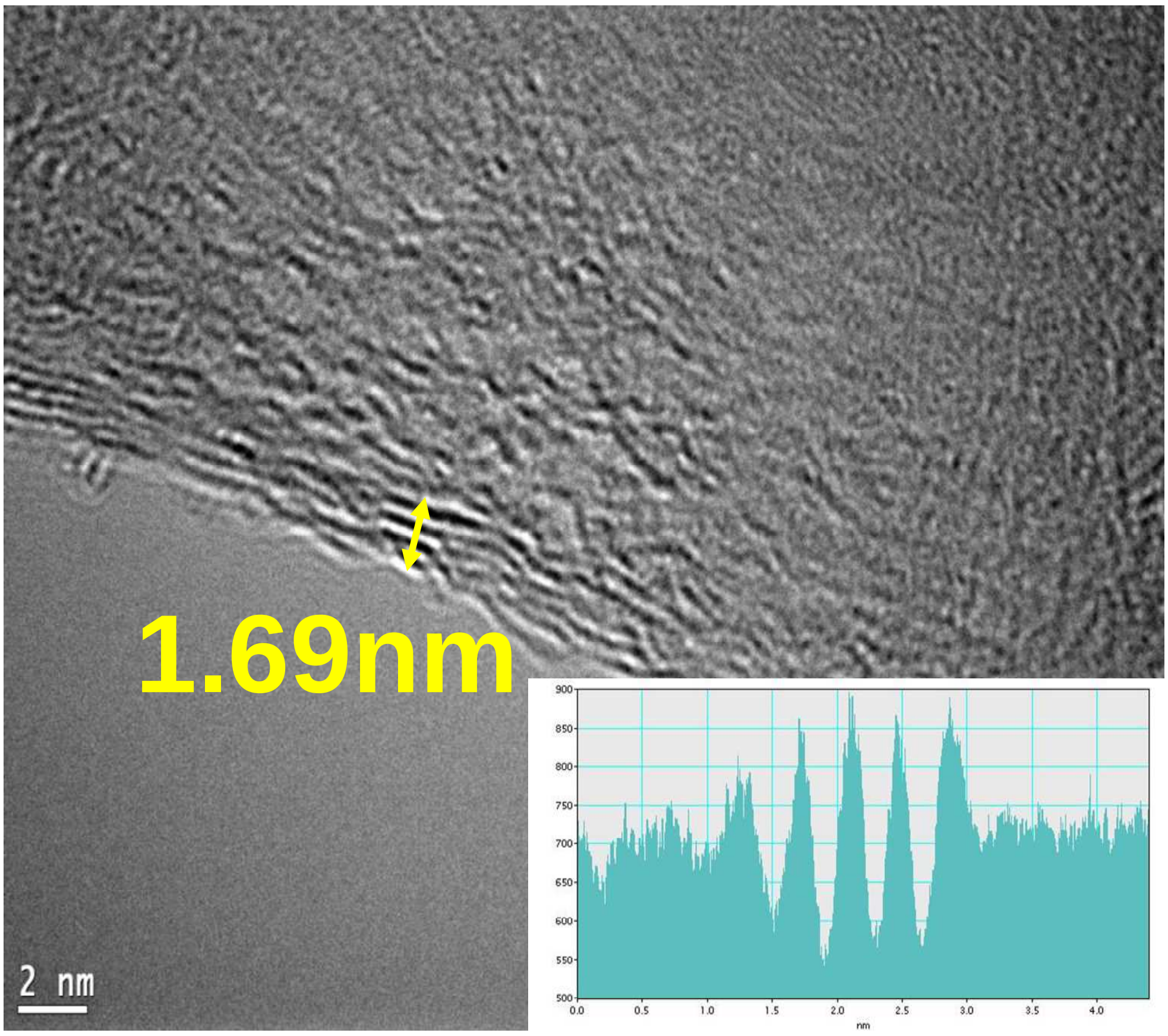

**Figure S1.** TEM image of the 4-layered GO with the corresponding cross-sectional profile (bottom inset): the arrow indicate the total thickness of 4-layers (1.69 nm), implicating the sheet separation of 0.42 nm.

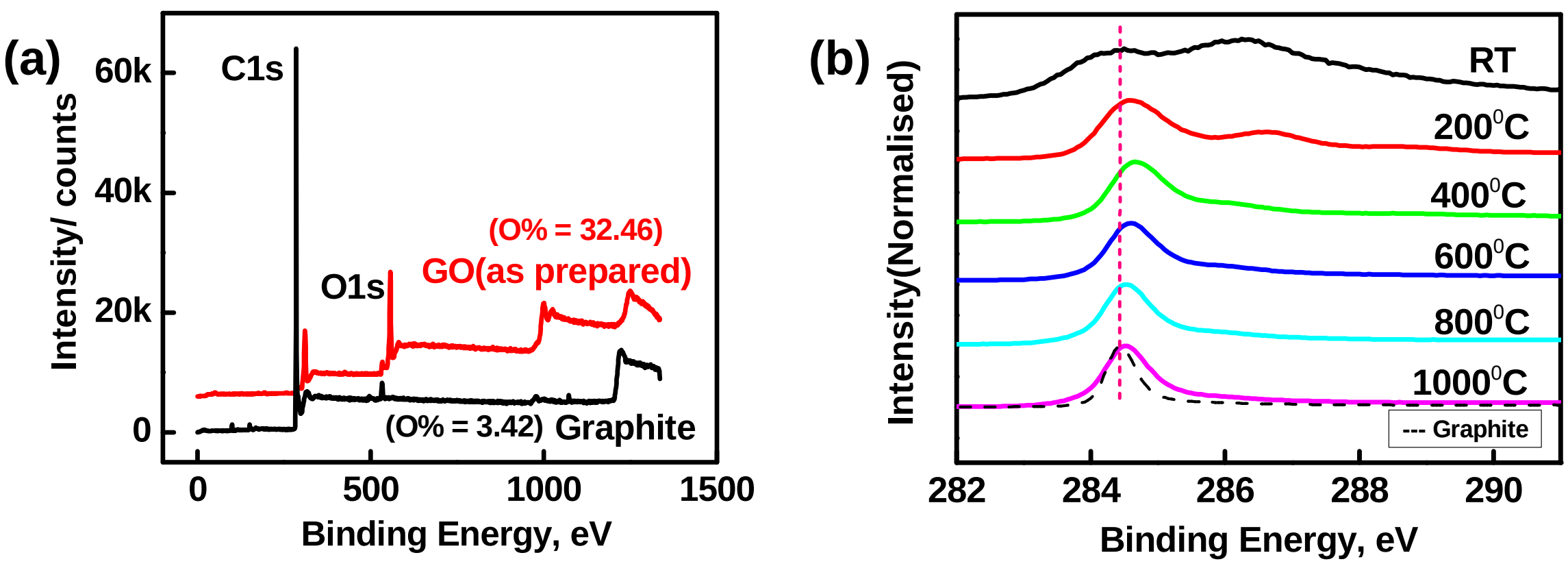

**Figure S2.** (a) WESS shifted due to charging GO. (b) High-resolution XPS spectra: C1s variation with temperature. The spectra were shifted in y-scale for clarity.

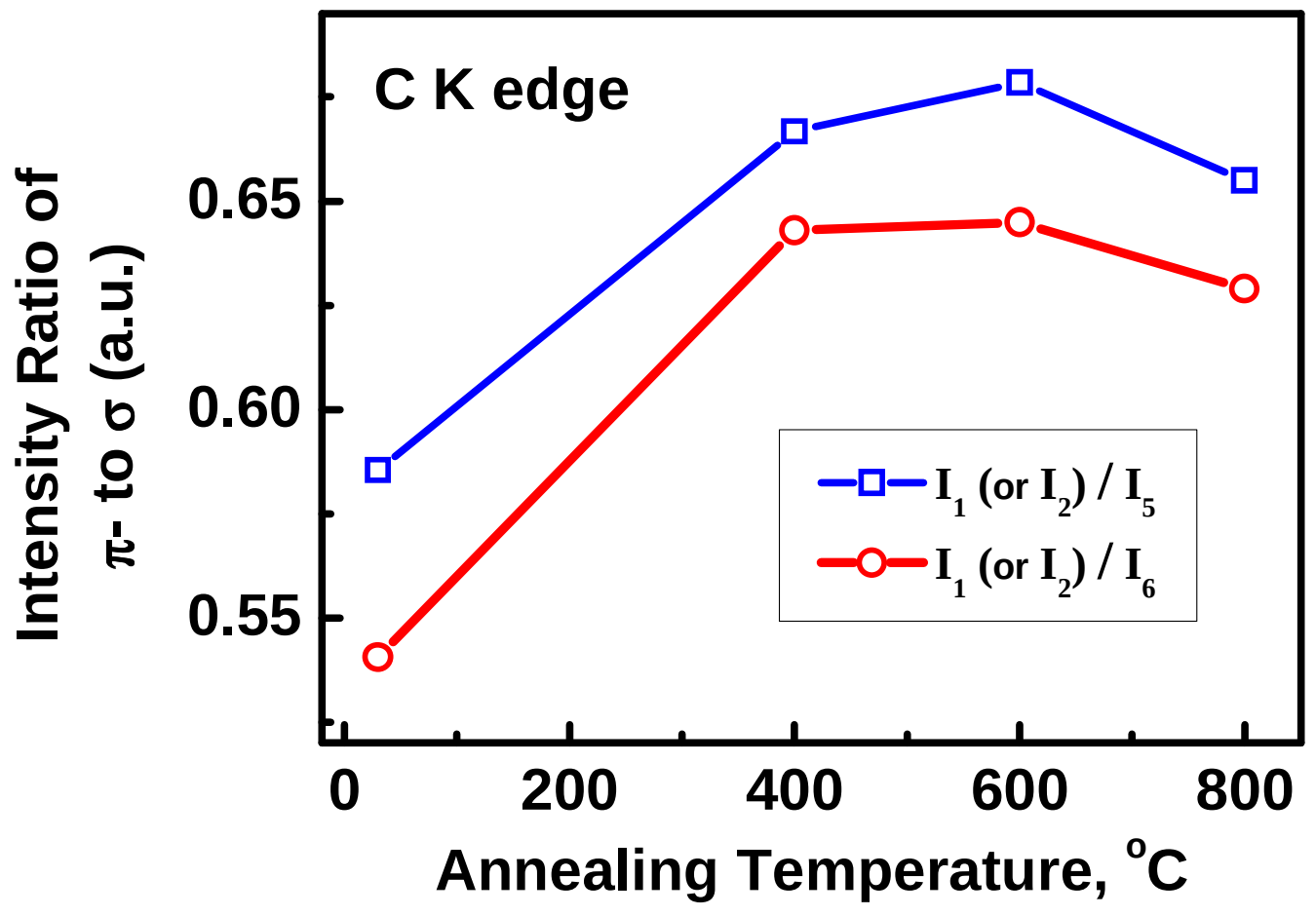

**Figure S3.** Intensity ratio of π* contribution (peak $I_1$ or $I_2$) to σ* (peak $I_5$ or $I_6$) as a function of temperature.